\pdfoutput=1

\documentclass[11pt,twoside,a4paper,cmspaper,final,collab]{cms-tdr}
\def\svnVersion{361967}\def\cmsCernNoTag{CERN-EP/2016-090}\def\cmsCernDate{\today}\def\cmsMessage{Published in Physical Review Letters as \href{http://dx.doi.org/10.1103/PhysRevLett.117.031802}{\doi{10.1103/PhysRevLett.117.031802.}}}

\begin{document}\cmsNoteHeader{EXO-14-005}

\hyphenation{had-ron-i-za-tion}
\hyphenation{cal-or-i-me-ter}
\hyphenation{de-vices}
\RCS$Revision: 353373 $
\RCS$HeadURL: svn+ssh://svn.cern.ch/reps/tdr2/papers/EXO-14-005/trunk/EXO-14-005.tex $
\RCS$Id: EXO-14-005.tex 353373 2016-07-06 07:31:50Z santanas $
\newlength\cmsFigWidth
\ifthenelse{\boolean{cms@external}}{\setlength\cmsFigWidth{1\columnwidth}}{\setlength\cmsFigWidth{0.7\textwidth}}
\ifthenelse{\boolean{cms@external}}{\providecommand{\cmsLeft}{top}}{\providecommand{\cmsLeft}{left}}
\ifthenelse{\boolean{cms@external}}{\providecommand{\cmsRight}{bottom}}{\providecommand{\cmsRight}{right}}
\ifthenelse{\boolean{cms@external}}{\providecommand{\CLp}{C.L.\xspace}}{\providecommand{\CLp}{CL\xspace}}

\cmsNoteHeader{EXO-14-005}
\title{Search for narrow resonances in dijet final states at \texorpdfstring{$\sqrt{s}=8$\TeV}{sqrt(s) = 8 TeV} with the novel CMS technique of data scouting}

\newcommand{\mjj}{\ensuremath{m_{\cmsSymbolFace{jj}}}\xspace}
\newcommand{\detajj}{\ensuremath{\abs{\Delta\eta_{\cmsSymbolFace{jj}}}}\xspace}
\newcommand{\Qstar}{\ensuremath{\PQ^*}\xspace}

\date{\today}

\abstract{A search for narrow resonances decaying into dijet final states is performed on data from proton-proton collisions at a center-of-mass energy of 8\TeV, corresponding to an integrated luminosity of 18.8\fbinv. The data were collected with the CMS detector using a novel technique called data scouting, in which the information associated with these selected events is much reduced, permitting collection of larger data samples. This technique enables CMS to record events containing jets at a rate of 1\unit{kHz}, by collecting the data from the high-level-trigger system.
In this way, the sensitivity to low-mass resonances is increased significantly, allowing previously inaccessible couplings of new resonances to quarks and gluons to be probed. The resulting dijet mass distribution yields no evidence of narrow resonances. Upper limits are presented on the resonance cross sections as a function of mass, and compared with a variety of models predicting narrow resonances. The limits are translated into upper limits on the coupling of a leptophobic resonance $\PZpr_B$ to quarks, improving on the results obtained by previous experiments for the mass range from 500 to 800\GeV.}

\hypersetup{%
pdfauthor={CMS Collaboration},%
pdftitle={Search for narrow resonances in dijet final states at sqrt(s)=8 TeV with the novel CMS technique of data scouting},%
pdfsubject={CMS},%
pdfkeywords={CMS, physics, exotica, resonances, dijet, CMS data scouting, novel technique}}

\maketitle

Deep inelastic proton-proton ($\Pp\Pp$) collisions often result in the production of two or more jets of particles, emitted with large transverse momentum \pt relative to the incoming beams. Within the framework of the standard model, the distribution of the invariant mass of the pair of jets having the largest values of \pt (the dijet) is expected to decrease smoothly with increasing mass. Many proposed extensions of the standard model predict the existence of new states coupling to quarks and gluons, which would manifest themselves as resonances in the dijet mass spectrum. The first results on this topic were presented by the UA1~\cite{Albajar:1988rs} and UA2~\cite{Alitti:1990kw,Alitti:1993pn} experiments, collecting data at the $\mathrm{Sp\bar{p}S}$ at a center-of-mass energy of 630\GeV. These results were later extended to larger values of resonance masses, by the Fermilab experiments CDF~\cite{Abe:1989gz,Abe:1993it,Abe:1995jz,Abe:1997hm,Aaltonen:2008dn} and D0~\cite{Abazov:2003tj} with $\sqrt{s}=1.8$ and  1.96\TeV proton-antiproton ($\Pp\PAp$) collisions at the Tevatron, and then by the CERN experiments ATLAS~\cite{Aad:2010ae,Aad:2011aj,Aad:2011fq,ATLAS:2012pu,Aad:2014aqa} and CMS~\cite{Khachatryan:2010jd,Chatrchyan:2011ns,CMS:2012yf,Chatrchyan:2013qha,Khachatryan:2015sja} with $\sqrt{s} =7$ and $8$\TeV $\Pp\Pp$ collisions at the LHC. Searches for dijet resonances with mass above $\sim$1.2\TeV were recently performed by ATLAS~\cite{ATLAS:2015nsi} and CMS~\cite{Khachatryan:2015dcf} using $\Pp\Pp$ collisions at $\sqrt{s}=13$\TeV. No significant excess was found in any of these searches. A review of the results below 8~\TeV can be found in Ref.~\cite{Harris:2011bh}.

Results obtained at different collider energies can be compared by translating the upper limits on the quark-quark resonance cross sections into upper bounds on the coupling constants $g$ of the new resonance to a pair of partons for a given model, as reported in Ref.~\cite{Dobrescu:2013cmh}. That study shows that these searches are not sensitive to the presence of low-mass ($\lesssim$1\TeV) resonances with small couplings ($g \lesssim 1$) to quarks. Similar conclusions can be obtained for quark-gluon and gluon-gluon resonances. The main experimental difficulties originate from the large cross section of multijet events at low dijet mass and the limited resources for processing and storing this data. Thus, the experiments are forced to increase the thresholds of their triggers, in order to keep an acceptable data volume despite the increasing collision rate.
Since the ultimate technical limitation is the available bandwidth at which events can be recorded on disk, an alternative strategy is reducing the event size in order to increase the recorded event rate. This approach, first implemented at the LHC by the CMS experiment in 2011, is referred to as {\textit{data scouting}~\cite{CMS:2012ooa}. A similar data collection strategy has also been recently introduced by the LHCb experiment \cite{LHCb-TurboStream}.

This Letter presents a search for narrow resonances decaying into two jets using pp interactions at $\sqrt{s}=8$\TeV. The study is performed on the scouting data collected with the CMS detector during 2012. The data set corresponds to an integrated luminosity of 18.8\fbinv.

The analysis strategy follows a similar search performed on the standard data streams~\cite{Khachatryan:2015sja}, which is only sensitive to resonance masses above 1200\GeV. Using data scouting, the present study extends the results of Ref.~\cite{Khachatryan:2015sja} down to 500\GeV.

The central feature of the CMS apparatus is a superconducting solenoid of 6\unit{m} internal diameter, providing a magnetic field of 3.8\unit{T}. Within the solenoid volume are a silicon pixel and strip tracker, a lead tungstate crystal electromagnetic calorimeter, and a brass and scintillator hadron calorimeters, each composed of a barrel and two endcap sections. Forward calorimeters extend the pseudorapidity ($\eta$) coverage provided by the barrel and endcap detectors. Muons are measured in gas-ionization detectors embedded in the steel flux-return yoke outside the solenoid. A more detailed description of the CMS detector, its coordinate system, and the main kinematic variables used in the analysis can be found in Ref.~\cite{Chatrchyan:2008aa}.

Like events in the standard data stream, data scouting events selected by the Level-1 (L1) trigger are processed online in the high-level trigger (HLT) computer farm. Jets are reconstructed from the energy deposits in the calorimeters. A momentum value is assigned to every calorimeter tower, with the magnitude equal to the measured energy and the direction given by the line from the center of the CMS detector to the center of the tower. Using the {\FASTJET} software~\cite{Cacciari:2011ma}, these momenta are clustered into jets with the anti-\kt algorithm~\cite{Cacciari:2008gp} with a size parameter of 0.5. The HLT selection requires the scalar sum of the jet transverse momenta (\pt) to be $\sum_\text{jet} \pt > 250$\GeV, where only jets with $\pt>40$\GeV and $\abs{\eta}<3$ are considered in the sum. This selection criterion translates into an accepted event rate of about 1\unit{kHz} at the HLT, at the highest instantaneous luminosity of 7.7\unit{Hz/nb} reached by the LHC in 2012. Scouting data are stored immediately after this selection. This event rate is comparable with the total allocated rate for the rest of the CMS physics program in the standard data stream. In order to sustain such a high rate in the data scouting stream, the amount of information stored for each of these HLT events is reduced to about 10\unit{kB}, consisting mainly of the four-momenta of the calorimeter jets reconstructed online,  compared with an event size of about 500\unit{kB} for normal data taking. Thus the 1\unit{kHz} of data scouting events consumes only about 2\% of the resources.
Since no raw data are stored in the data scouting stream, it is not possible to perform an offline reconstruction of these events.

The scouting events are analyzed using standard CMS software tools. Jet momenta are corrected using the calibration constants derived from simulations, test beam results, and $\Pp\Pp$ collision data~\cite{Chatrchyan:2011ds}. The energy density $\rho$ in the event in a unit of area $\Delta \eta{\times}\Delta \phi$ is also stored in the scouting record. Energies from additional collisions in the same or adjacent bunch crossings (pileup) are subtracted from the jet energies using an event-by-event and jet-area based correction~\cite{jetarea_fastjet_pu}. All jets in this analysis are required to have $\pt>30$\GeV and $\abs{\eta}<2.5$ and to pass a jet quality selection, based on the transverse distribution and magnitude of the energy deposits in the electromagnetic and hadron calorimeters. Additional requirements are applied to remove events where electronic noise mimics jets. Finally, the azimuthal angle between the two highest \pt jets is required to be $\Delta \phi > \pi/3$. The latter requirement enforces the dijet event topology and suppresses the instrumental background originating in the calorimeters.

As in previous CMS dijet searches~\cite{Chatrchyan:2011ns,CMS:2012yf,Chatrchyan:2013qha,Khachatryan:2015sja}, geometrically close jets are combined into ``wide jets''.  In this process, the two leading jets are used as seeds and the four-momenta of all other jets are then added to the closest leading jet, if they satisfy the condition $\Delta R=\sqrt{\smash[b]{(\Delta\eta)^2 + (\Delta\phi)^2}}<1.1$, to obtain two wide jets, which then form the dijet system. By recovering the final state radiation from quarks and gluons, the wide-jet algorithm improves the dijet mass resolution.

The background from $t$-channel dijet events is suppressed by requiring the pseudorapidity separation of the two wide jets to satisfy $\detajj<1.3$. This requirement maximizes the search sensitivity for isotropic decays of dijet resonances in the presence of QCD multijet background.
We select events with $\mjj > 390\GeV$, for which the combined L1 trigger and HLT are found to be fully efficient.

Figure~\ref{fig:bkgFitData} shows the measured differential cross section as a function of dijet mass in bins of variable width corresponding to the dijet mass resolution~\cite{Khachatryan:2010jd}. The plot is limited to the mass range relevant for this study. To test the smoothness of the measured cross section, a binned maximum likelihood fit to the data is performed with the parametrization:
\begin{equation} \label{eq:bkgShape}
\frac{{\rd}\sigma}{{\rd}\mjj} =
\frac{P_{0} (1 - x)^{P_{1}}}{x^{P_{2} + P_{3} \ln{(x)}}},
\end{equation}
where $x=\mjj/\sqrt{s}$ and $P_i$ ($i=0$, 1, 2, 3) are free parameters. This functional form was used previously to describe both data and QCD predictions~\cite{Aaltonen:2008dn,Aad:2010ae,Aad:2011aj,Aad:2011fq,ATLAS:2012pu,Aad:2014aqa,Khachatryan:2010jd,Chatrchyan:2011ns,CMS:2012yf,Chatrchyan:2013qha,Khachatryan:2015sja}.  The fit to data yields a $\chi^2$ of 31 for 22 degrees of freedom. The difference between the data and the best fit function is shown at the bottom of Fig.~\ref{fig:bkgFitData}, normalized to the statistical uncertainty of the data in each bin. No significant excess of events above the background fit is observed.

\begin{figure}
\centering
\includegraphics[width=\cmsFigWidth]{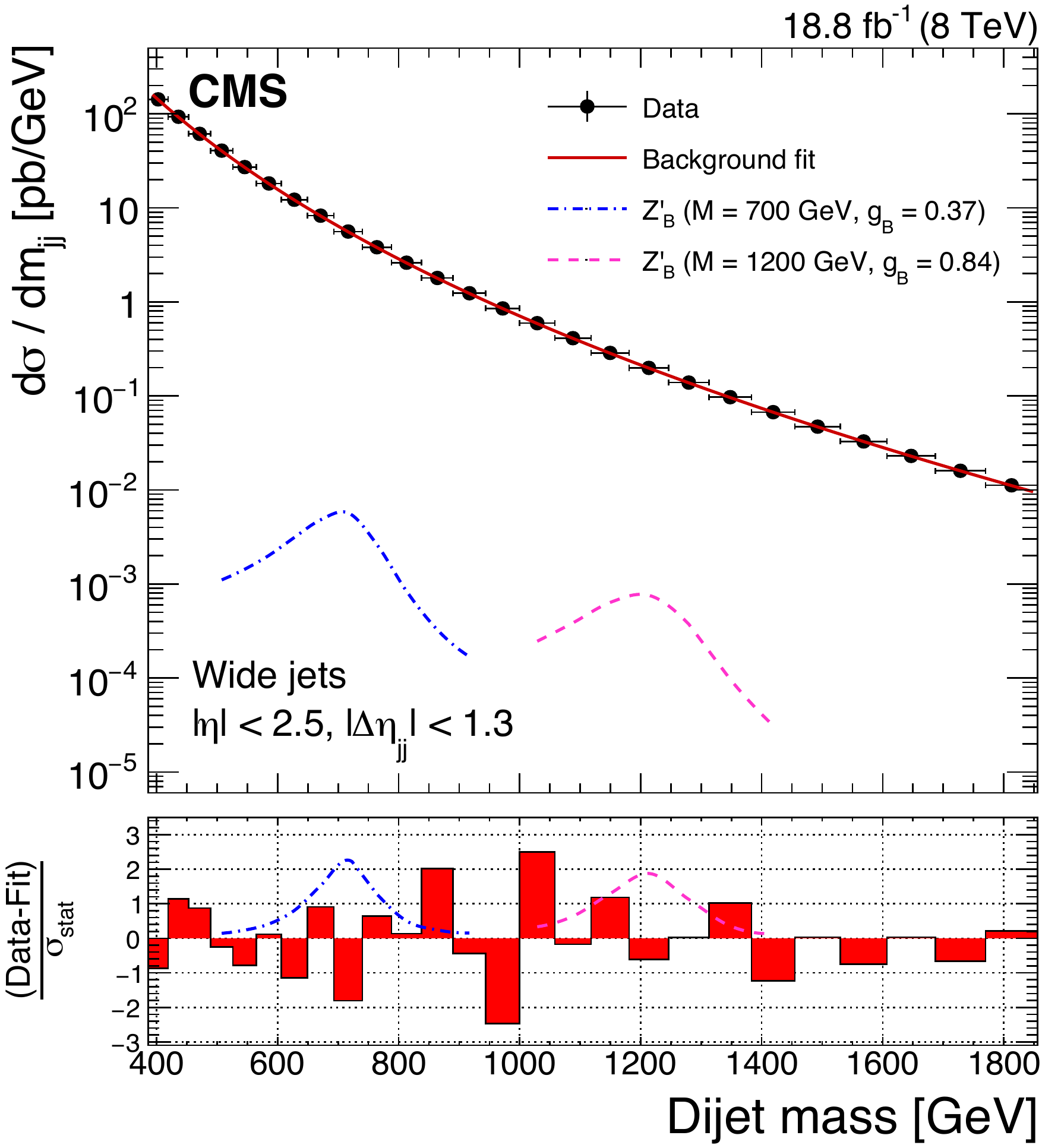}
\caption{ The reconstructed dijet mass distribution (black points) fitted with the function of Eq. (\ref{eq:bkgShape}) (red solid line).
The bottom panel compares the data and the fit result, normalized by the statistical uncertainty in the data, for each bin. The predicted distributions of narrow resonance signals for a hypothetical leptophobic resonance $\PZpr_B$ \cite{PhysRevD.82.011901, PhysRevLett.110.231801} with two different mass and coupling values are shown in both panels (dash-dotted curves). This dijet mass distribution complements that observed at higher mass~\cite{Khachatryan:2015sja}.\label{fig:bkgFitData}}
\end{figure}

This search focuses on narrow resonances with intrinsic widths smaller than the dijet mass resolution.
Three varieties of narrow resonances associated with representative simplified models are considered: Randall--Sundrum (RS) gravitons~\cite{ref_rsg} coupling either to quark-antiquark ($\PQq\PAQq$) or to gluon-gluon ($\Pg\Pg$) pairs with coupling $k/\Mpl = 0.1$, and excited quarks~\cite{ref_qstar,Baur:1989kv} coupling to quark-gluon ($\PQq\Pg$) pairs with compositeness scales that are equal to the excited quark masses.
The distribution of the dijet mass for signal events is modeled using the {\PYTHIA 8.2}~\cite{Sjostrand:2007gs} generator with tune 4C~\cite{Corke:2010yf} for the description of the underlying event. The generated events are processed through a \GEANTfour~\cite{refGEANT} model of the CMS detector. The jet energies in the simulated signal samples are corrected in order to match the energy scale observed in the online reconstruction in the following way: a small fraction of the scouting data, corresponding approximately to one million events, is also saved in the standard data record, to allow a jet-by-jet comparison of the offline and online reconstruction performances. The corrections are derived from this subset of the data sample by comparing the energy of jets reconstructed offline with the energy of the corresponding jets reconstructed at the HLT, calculated as a function of \pt and $\eta$ of the offline reconstructed jet. The energies of online and offline reconstructed jets agree within $\sim$2\% in the kinematic phase-space of this analysis. Figure~\ref{fig:sigMjj} shows the $\PQq\PQq$, $\PQq\Pg$, and $\Pg\Pg$ signal shapes for a resonance with mass of 900\GeV. The predicted mass distributions have Gaussian cores due to both QCD radiation and jet energy resolution and tails towards lower mass values primarily from QCD radiation.
The contribution of this low-mass tail to the line shape depends on the parton content of the resonance. Resonances containing gluons, which emit QCD radiation more strongly
than quarks, have a more pronounced tail. Neglecting the tails, the approximate width of the reconstructed dijet mass peak varies with resonance mass
from 11\% at 500\GeV to 7\% at 1600\GeV.
\begin{figure}[!hbt]
\centering
\includegraphics[width=\cmsFigWidth]{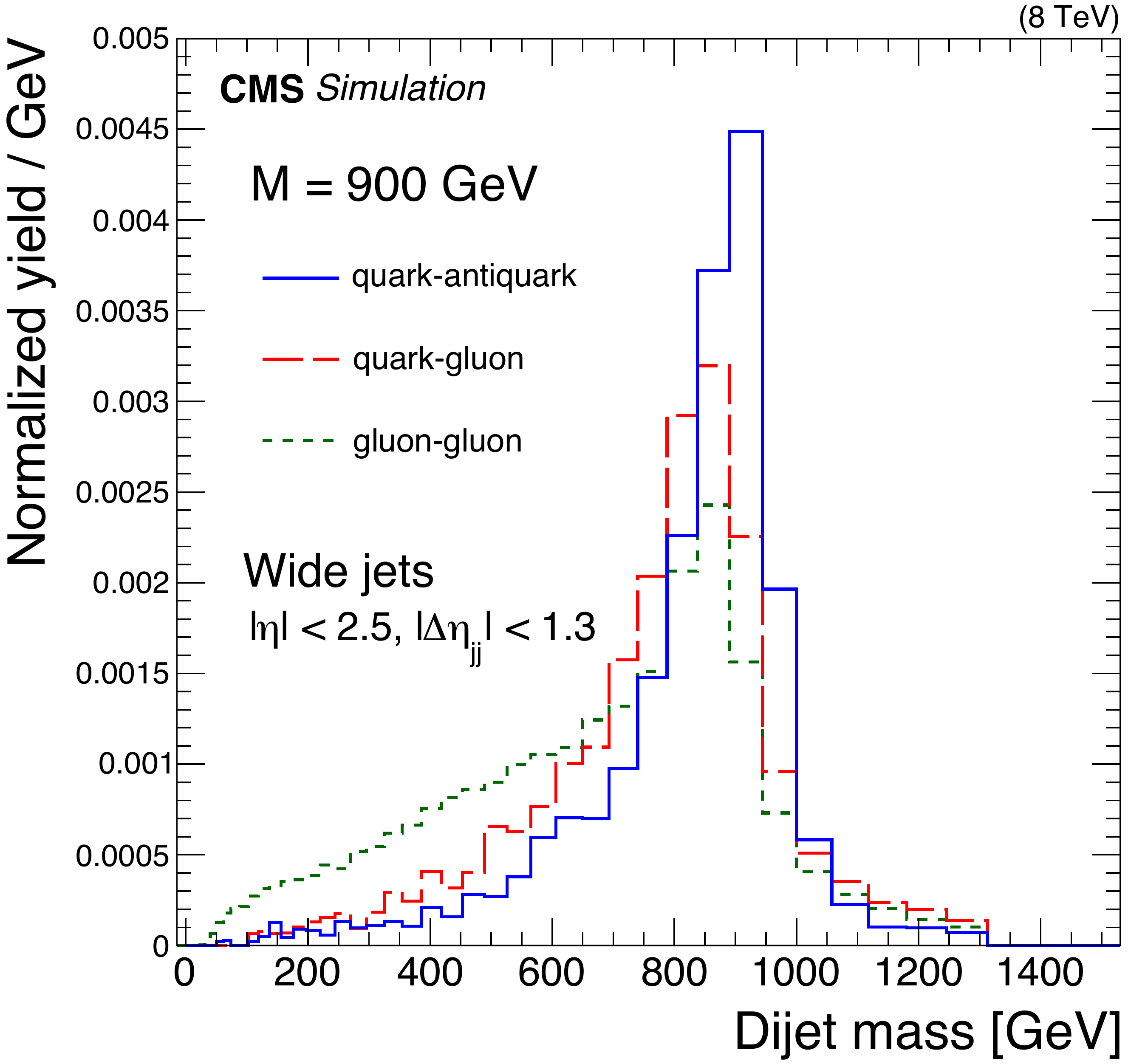}
\caption{ The reconstructed dijet mass distributions for simulated RS gravitons decaying to quark-antiquark, excited quarks decaying to quark-gluon, and  RS gravitons decaying to gluon-gluon, for a resonance mass of 900\GeV.
\label{fig:sigMjj}}
\end{figure}

A Bayesian analysis~\cite{statistics_review} is used to set upper limits on the signal cross section. A uniform prior for the positive signal cross section is assumed. We calculate the posterior probability density as a function of the resonance cross section for resonance masses between 500 and 1600\GeV, in 100\GeV steps.
At each considered mass value, a signal-plus-background fit to the data is performed, with the signal production cross section left as a free parameter. The contribution of the background is taken to be the distribution given by the background component of the fitted function.
The likelihood is formed using the following inputs: the reconstructed dijet mass distribution in data, the background estimate from the best fit of the background hypothesis to the data, and the simulated \mjj distribution for the resonance that is a histogram normalized to the resonance cross section.

The dominant sources of systematic uncertainty are the jet energy scale and resolution, the luminosity, and the estimation of the background.
The uncertainty of 5\% in the jet energy scale is derived from the results reported in Ref.~\cite{Chatrchyan:2011ds} and from dedicated studies performed for the data scouting analysis.
This uncertainty is propagated to the limits by scaling the dijet signal distribution by 5\% up and down. The observed limits change by few percent when varying the jet energy scale uncertainty from 5\% to 1\%.
The uncertainty in the jet energy resolution translates into an uncertainty of $\pm$10\% in the resolution of the signal distribution~\cite{Chatrchyan:2011ds}, and is propagated to the limits by increasing and decreasing the width of the simulated dijet mass distribution shape for the signal by 10\%. The uncertainty in the normalization of the signal resulting from the uncertainty in the integrated luminosity is $\pm$2.6\%~\cite{CMS-PAS-LUM-13-001}. Uncertainties in the values of the parameters describing the background also contribute to the uncertainty in the signal strength and are taken into account in the limit setting procedure, as described below.

The above systematic uncertainties are incorporated in the limit calculation as nuisance parameters. Log-normal priors are used to model the jet energy scale, jet energy resolution, and luminosity uncertainties, while uniform priors are used for the parameters of the background function. The posterior function for the signal cross section is obtained by marginalizing over these nuisance parameters. The integration is performed independently for each of the background nuisance parameters in a range around the best-fit values that is large enough to accommodate a decrease in the likelihood by a factor of 1000 from its maximum value.

Figure~\ref{fig:xsecUL} shows the observed model-independent upper limits at 95\% confidence level (\CLp) on $\sigma \mathcal{B} A$, \ie the product of the signal cross section ($\sigma$), the branching fraction to jets ($\mathcal{B}$), and the acceptance ($A$) for the kinematic requirements $\detajj<1.3$ and $\abs{\eta}<2.5$. The effect of the $m_{jj}>390\GeV$ requirement has been taken into account by correcting the limits, and therefore does not appear in the acceptance. In Table~\ref{limits-table}, the limits are reported separately for narrow $\PQq\PQq$, $\PQq\Pg$, and $\Pg\Pg$ resonances, in the mass range between 500 and 1600\GeV. The data scouting exclusions for masses between 1200 and 1600\GeV overlap with those from the high-mass search performed with the standard data stream~\cite{Khachatryan:2015sja}. In the overlapping mass interval, the sensitivity of the data scouting search is found to be almost identical with the high-mass search sensitivity, and the results are found to be compatible with those of the high-mass search within $\pm$2 standard deviations.

Additionally, we apply our search results to the following models of $s$-channel dijet resonances: excited quarks~\cite{ref_qstar,Baur:1989kv}; axigluons~\cite{ref_axi,Chivukula:2013xla}; scalar diquarks~\cite{ref_diquark}; new gauge bosons ($\PWpr$ and $\PZpr$)~\cite{ref_gauge}; and RS gravitons~\cite{ref_rsg}. More details on the specific choices of couplings for the models considered can be found in Ref.~\cite{CMS:2012yf}.  The upper limits presented are compared to the parton-level predictions of $\sigma \mathcal{B} A$ without any detector simulation, in order to determine mass limits on new particles. The model predictions shown in Fig.~\ref{fig:xsecUL} are calculated in the narrow-width approximation~\cite{Harris:2011bh} using CTEQ6L1 PDFs ~\cite{refCTEQ} at leading order and a next-to-leading-order k-factor is included for the $\PWpr$, $\PZpr$, and axigluon models~\cite{Chivukula:2013xla}.
New particles are excluded at 95\% \CLp~in mass regions for which the theoretical curve lies at or above the observed upper limit for the appropriate final state in Fig.~\ref{fig:xsecUL}. The RS graviton cross section can be compared to the average of the upper limits for $\PQq\PQq$ and $\Pg\Pg$ final states.

\begin{table}
\centering
\topcaption{Observed upper limits at 95\% CL on $\sigma \mathcal{B}A$ for resonances decaying to $\PQq\PQq$, $\PQq\Pg$, and $\Pg\Pg$ final states as a function of the resonance mass.}
\label{limits-table}
\newcolumntype{.}{D{.}{.}{2.2}}
\begin{scotch}{c...}
\multirow{2}{*}{Mass [\GeVns{}] }&\multicolumn{3}{c}{Upper limit on $\sigma \mathcal{B}A$ [pb]} \\ \cline{2-4}
  & \multicolumn{1}{c}{$\PQq\PQq$}  & \multicolumn{1}{c}{$\PQq\Pg$}  & \multicolumn{1}{c}{$\Pg\Pg$} \\
\hline
500             & 3.0   & 11   & 11 \\
600            & 1.3   & 3.4  & 3.7\\
700            & 0.66   & 1.1  & 1.5\\
800            & 1.0  & 1.5 & 2.1 \\
900            & 0.50  & 1.4  & 2.2\\
1000          & 0.44  & 0.65 & 0.85\\
1100           & 0.40  & 0.66 & 1.0\\
1200          & 0.25  & 0.49 & 0.70\\
1300          & 0.19  & 0.30 & 0.40\\
1400          & 0.10  & 0.19 & 0.29\\
1500          & 0.07  & 0.10 & 0.14\\
1600          & 0.05 & 0.07 & 0.10\\
\end{scotch}
\end{table}

We exclude excited quarks, axigluons, scalar diquarks, $\PWpr$ and $\PZpr$ bosons, and RS gravitons with masses between 500 and 1600\GeV. These limits cover regions not excluded by previous dijet resonance searches at hadron colliders, as follows: scalar diquarks with masses between 630 and 970\GeV~\cite{Harris:2011bh}; $\PWpr$ bosons with masses between 840 and 1000\GeV~\cite{Harris:2011bh}; $\PZpr$ bosons with masses between 740 and 1000\GeV~\cite{Harris:2011bh,CMS:2012yf}; RS gravitons with masses between 500 and 1000\GeV~\cite{Harris:2011bh}.

\begin{figure}
\centering
\includegraphics[width=\cmsFigWidth]{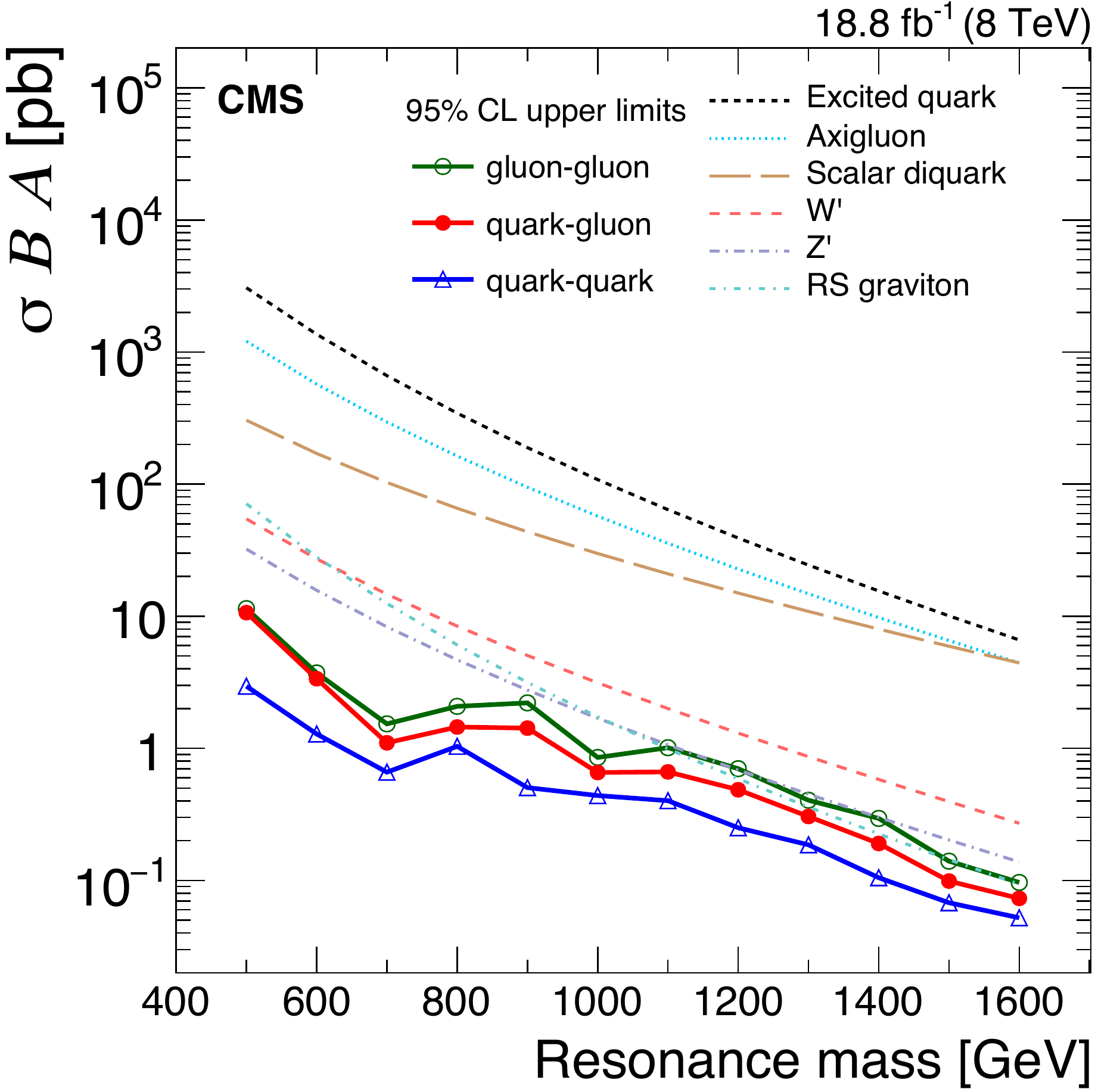}
\caption{Observed 95\% \CLp~upper limits on $\sigma \mathcal{B}A$ for a narrow resonance decaying to gluon-gluon final states (open circles), quark-gluon final states (solid circles), and quark-quark final states (open triangles) compared with theoretical predictions for various resonance models.
\label{fig:xsecUL}}
\end{figure}

Following the theoretical framework of Ref.~\cite{Dobrescu:2013cmh}, the model-independent upper limits on the cross section of $\PQq\PQq$ resonances are translated into 95\% \CLp~upper limits on the coupling $g_B$ of a hypothetical leptophobic resonance $\PZpr_B \to\Pq \PAQq$ as a function of its mass.  The $\PZpr_B$ production cross section scales with the square of the coupling $g_B$.  Figure~\ref{fig:gBvsMZ} shows the upper limits obtained with the data scouting technique in the mass region from 500 to 1200\GeV, extending the coverage of previous CMS searches to below 1200\GeV.  Previous exclusions obtained with similar searches at various collider energies are also shown. As a result of the large data set collected by the data scouting stream, the bound on $g_B$ is improved by up to a factor of 3 for resonance masses between 500 and 800\GeV, compared to previous searches. This corresponds to an order-of-magnitude improvement in the cross section limit.

\begin{figure}
\centering
\includegraphics[width=\cmsFigWidth]{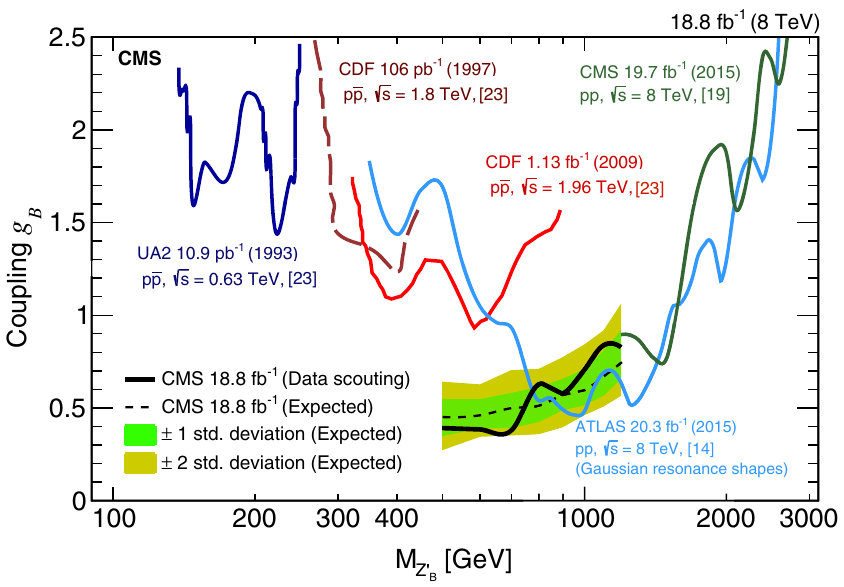}
\caption{Observed 95\% \CLp~upper limits on the coupling $g_B$ of a hypothetical leptophobic
resonance $\PZpr_B \to\PQq \PAQq$~\cite{Dobrescu:2013cmh} as a function of its mass. The results from this study are compared to results obtained with similar searches at different collider energies~\cite{Dobrescu:2013cmh, Aad:2014aqa}.\label{fig:gBvsMZ}}
\end{figure}

In summary, a search for narrow resonances decaying into two jets was performed using data from proton-proton collisions recorded by the CMS experiment at $\sqrt{s}=8$\TeV, corresponding to an integrated luminosity of 18.8\fbinv. The novel technique of data scouting was used; by reducing the information stored per event, multijet events could be collected in sufficiently large samples that a sensitive search for  dijet resonances down to masses as low as 500\GeV was possible. No evidence for a narrow resonance is found. Model-independent upper limits on production cross sections are derived for quark-quark, quark-gluon, and gluon-gluon resonances. Based on these results, new limits are set on an extensive selection of narrow $s$-channel resonances over mass ranges not excluded by previous searches at hadron colliders. Bounds on the coupling of a hypothetical leptophobic resonance decaying to quark-antiquark are also provided, as a function of the resonance mass. The limits obtained are the most stringent to date in the dijet final state for narrow resonance masses between about 500 and 800\GeV.

\begin{acknowledgments}
We congratulate our colleagues in the CERN accelerator departments for the excellent performance of the LHC and thank the technical and administrative staffs at CERN and at other CMS institutes for their contributions to the success of the CMS effort. In addition, we gratefully acknowledge the computing centers and personnel of the Worldwide LHC Computing Grid for delivering so effectively the computing infrastructure essential to our analyses. Finally, we acknowledge the enduring support for the construction and operation of the LHC and the CMS detector provided by the following funding agencies: BMWFW and FWF (Austria); FNRS and FWO (Belgium); CNPq, CAPES, FAPERJ, and FAPESP (Brazil); MES (Bulgaria); CERN; CAS, MoST, and NSFC (China); COLCIENCIAS (Colombia); MSES and CSF (Croatia); RPF (Cyprus); MoER, ERC IUT and ERDF (Estonia); Academy of Finland, MEC, and HIP (Finland); CEA and CNRS/IN2P3 (France); BMBF, DFG, and HGF (Germany); GSRT (Greece); OTKA and NIH (Hungary); DAE and DST (India); IPM (Iran); SFI (Ireland); INFN (Italy); MSIP and NRF (Republic of Korea); LAS (Lithuania); MOE and UM (Malaysia); BUAP, CINVESTAV, CONACYT, LNS, SEP, and UASLP-FAI (Mexico); MBIE (New Zealand); PAEC (Pakistan); MSHE and NSC (Poland); FCT (Portugal); JINR (Dubna); MON, RosAtom, RAS and RFBR (Russia); MESTD (Serbia); SEIDI and CPAN (Spain); Swiss Funding Agencies (Switzerland); MST (Taipei); ThEPCenter, IPST, STAR and NSTDA (Thailand); TUBITAK and TAEK (Turkey); NASU and SFFR (Ukraine); STFC (United Kingdom); DOE and NSF (USA).
\end{acknowledgments}
\bibliography{auto_generated}

\cleardoublepage \appendix\section{The CMS Collaboration \label{app:collab}}\begin{sloppypar}\hyphenpenalty=5000\widowpenalty=500\clubpenalty=5000\textbf{Yerevan Physics Institute,  Yerevan,  Armenia}\\*[0pt]
V.~Khachatryan, A.M.~Sirunyan, A.~Tumasyan
\vskip\cmsinstskip
\textbf{Institut f\"{u}r Hochenergiephysik der OeAW,  Wien,  Austria}\\*[0pt]
W.~Adam, E.~Asilar, T.~Bergauer, J.~Brandstetter, E.~Brondolin, M.~Dragicevic, J.~Er\"{o}, M.~Flechl, M.~Friedl, R.~Fr\"{u}hwirth\cmsAuthorMark{1}, V.M.~Ghete, C.~Hartl, N.~H\"{o}rmann, J.~Hrubec, M.~Jeitler\cmsAuthorMark{1}, A.~K\"{o}nig, I.~Kr\"{a}tschmer, D.~Liko, T.~Matsushita, I.~Mikulec, D.~Rabady, N.~Rad, B.~Rahbaran, H.~Rohringer, J.~Schieck\cmsAuthorMark{1}, J.~Strauss, W.~Treberer-Treberspurg, W.~Waltenberger, C.-E.~Wulz\cmsAuthorMark{1}
% ...
\vskip\cmsinstskip
% ...
1:~~Also at Vienna University of Technology, Vienna, Austria\\
% ...

\end{sloppypar}
\end{document}